\documentstyle[epsfig,here,12pt]{article}
\begin{document} 
\newcommand {\ber}
{\begin{eqnarray*}}
 \newcommand {\eer} {\end{eqnarray*}}
\newcommand {\bea}
{\begin{eqnarray}} \newcommand {\eea} {\end{eqnarray}} \newcommand {\beq}
{\begin{equation}} \newcommand {\eeq} {\end{equation}}
\newcommand {\state}
[1] {\mid \! \! {#1} \rangle} \newcommand {\eqref} [1] {(\ref {#1})}
\newcommand{\preprint}[1]{\begin{table}[t] 
           \begin{flushright}               
           \begin{large}{#1}\end{large}     
           \end{flushright}                 
           \end{table}}                     
\def\Acknowledgements{\bigskip  \bigskip {\begin{center} \begin{large}
             \bf ACKNOWLEDGEMENTS \end{large}\end{center}}}

\def\Dslash{\not{\hbox{\kern-4pt $D$}}}
\def\CR{\nonumber \\ }
\def\cmp#1{{\it Comm. Math. Phys.} {\bf #1}}
\def\cqg#1{{\it Class. Quantum Grav.} {\bf #1}}
\def\pl#1{{\it Phys. Lett.} {\bf #1B}}
\def\prl#1{{\it Phys. Rev. Lett.} {\bf #1}}
\def\prd#1{{\it Phys. Rev.} {\bf D#1}}
\def\prr#1{{\it Phys. Rev.} {\bf #1}}
\def\np#1{{\it Nucl. Phys.} {\bf B#1}}
\def\ncim#1{{\it Nuovo Cimento} {\bf #1}}
\def\jmath#1{{\it J. Math. Phys.} {\bf #1}}
\def\mpl#1{{\it Mod. Phys. Lett.}{\bf A#1}}
\def\jmp#1{{\it J. Mod. Phys.}{\bf A#1}}
\def\mycomm#1{\hfill\break{\tt #1}\hfill\break}
\begin{titlepage}
\titlepage
\rightline{{CPTH-S090.1100}}
\rightline{WIS-00/24/Nov-DPP}
\rightline{TAUP-2646-2000}

\vskip 1cm
\centerline{{\Large \bf Massless  ${\rm QCD}_2$ From Current Constituents}}
\vskip 1cm
\centerline{A. Armoni$^a$
, Y. Frishman$^b$
, J. Sonnenschein$^c$}
\begin{center}
\em $^a$Centre de Physique Th{\'e}orique de l'{\'E}cole 
Polytechnique
\\91128 Palaiseau Cedex, France
\end{center}
\begin{center}
\em $^b$Department of Particle Physics
\\Weizmann Institute of Science
\\76100 Rehovot, Israel
\end{center}
\begin{center}
\em $^c$School of Physics and Astronomy
\\Beverly and Raymond Sackler Faculty of Exact Sciences
\\Tel Aviv University, Ramat Aviv, 69978, Israel
\end{center}

\begin{abstract} 
We discuss the spectra of multi-flavor massless ${\rm QCD}_2$.
 An approximation in which the Hilbert space is truncated
 to two currents states is used.
We write down a 't Hooft like equation for the wave function of
 the two currents
states. We solve this equation for the lowest massive state and find an
excellent agreement with the DLCQ results.
 In addition, the 't Hooft model and the large $N_f$ limit
spectra are re-derived by using a description in terms of currents.
 
\end{abstract}
\end{titlepage}

\newpage
\section{Introduction}
Two dimensional quantum chromodynamics (${\rm QCD}_2$) is a useful toy
model
 for the real world QCD. The large $N_c$ model was solved by 't
 Hooft\cite{thooft}
and shows confinement of quarks with an approximately linear
Regge trajectory of states.
Other issues, such as the baryonic spectrum at strong coupling\cite{DFS}
and questions of screening versus confinement\cite{GKMS,AFS} can
 also be addressed in this framework.

The two-dimensional model with fermions in the adjoint representation
 is also interesting and attracted a lot of attention in recent
 years\cite{DK,kutasov,BDK,KS,AS,dalley,GHK,AP,uwe}.
In particular it was shown in \cite{KS} that the adjoint fermions model 
is equivalent to ${\rm QCD}_2$ model with level $N_f=N_c$, for the massive part
of the spectrum, in the case of massless fermions. 
Another attempt to address the adjoint fermions model, was by using the 
currents as building blocks of the spectrum \cite{AS}. 

The idea of the present work is to study the spectrum of  ${\rm QCD}_2$
at arbitrary level $N_f$ using states built from two currents,
for the case of massless fermions.

Our goal in this work is two folded:
(i) To derive a `t Hooft like equation for 
the wave function of the ``currentball" states 
 for ${\rm QCD}_2$ 
models at arbitrary level $N_f$. The equation should interpolate   
 between the description of a single
flavor model ('t Hooft model), the adjoint fermions model $N_f=N_c$ and
the large $N_f$ model.
(ii) To solve the equation for the lowest massive state.

Whereas the `t Hooft model $N_f=1$ is exactly solvable, 
the multi-flavor case $N_f>1$ is not  solvable model even in the
Veneziano
limit when both $N_c$ and $N_f$ are taken to infinity (with a fixed
ratio), since pair creation and annihilation is not suppressed. 
In the present work , we use an approximation in which we restrict
ourselves
to two currents states. We cannot justify a-priori such an
 approximation for arbitrary
level. However the numerical solutions for the lowest massive state 
admit a very close resemblance to the DLCQ results where such a 
truncation was not used \cite{GHK,AP,uwe}. A justification can be given for
 ${N_f \over N_c}$ very
small or very large. 

The obtained equation
 suggests the following picture: the underlying degrees of freedom in
the problem are interacting ``gluons'' with mass $e^2 N_f \over \pi$.
Actually, these are really quanta of the the color currents.
As it is well known, there are no independent gluon degrees
of freedom in two dimensions.
 
The organization of the manuscript is as follows. In section 2
 the bosonization of multi-flavor QCD is reviewed. In section 3 we state
the
problem of diagonalizing the mass operator in terms of currents and
arrive at a 't Hooft like equation for the currents wavefunction.
 Section 4 is devoted to a solution of the differential equation and
 for a discussion in some specific cases, such as the 't Hooft limit,
 the $N_f\gg N_c$ case and the adjoint fermions ($N_f=N_c$) case.
 Section 4 is a summary and a discussion.

\section{Massless ${\rm QCD}_2$ and Bosonization}
Massless multi-flavor ${\rm QCD}_2$ with fermions in the fundamental
representation of $SU(N_c)$
 is described by the following action
\beq
\label{qcd} S=\int d^2x \ {\rm tr}\ (-{1\over 2e^2}F_{\mu\nu}^2+i\bar \Psi\Dslash\Psi )
 \eeq
where $\Psi = \Psi ^i_a$, $i=1 \dots N_c$, $a=1 \dots N_f$.

It is natural to bosonize this system, since bosonization in the
$SU(N_c)\times SU(N_f)\times U_B(1)$ scheme decouples
color and flavor degrees of freedom (in the massless case).
The bosonized form of the action of this theory is given by\cite{FSreview}
\bea
\label{bosonized} \lefteqn{ S_{bosonized} =} \\
 && N_f S_{WZW}(h) + N_c S_{WZW}(g)  + \int d^2x\ {1\over 2}\partial _\mu \phi \partial ^\mu
 \phi -\int d^2x\ {\rm tr}\ {1\over {2e^2}} F
 _{\mu \nu} F^{ \mu \nu} 
\nonumber \\
 && -{N_f\over 2\pi} \int d^2 x \ {\rm tr}\ (ih^\dagger \partial_+ h A_-
+ih\partial_ - h^\dagger A_+ + A_+ h A_- h^\dagger - A_+ A_-)   \nonumber
\eea
where $h \in SU(N_c)$, $g\in SU(N_f)$, $\phi$ is the baryon number
 and $S_{WZW}$ stands for the
Wess-Zumino-Witten action, which for complex fermions reads
\ber
\lefteqn{S_{WZW}(g)={1\over{8\pi}}\int _\Sigma d^2x \ {\rm tr}\ (\partial _\mu
g\partial ^\mu g^{-1}) + } \\
 && {1\over{12\pi}}\int _B d^3y \epsilon ^{ijk} \
 {\rm tr}\ (g^{-1}\partial _i g) (g^{-1}\partial _j g)(g^{-1}\partial _k g),
\eer
Since we are interested in the massive spectrum of the theory and
the flavor degrees of freedom are entirely decoupled from the
system and they are massless, we can put aside the $g$ and $\phi$ fields (There
is a residual interaction of the zero modes of the $g,h$ and $\phi$ fields, but
it is not important to our discussion\cite{KS}).

Upon choosing the light cone gauge $A _- = 0$ and integrating $A_+$ we
arrive to the following action
\beq
S = N_f S_{WZW}(h) - {1\over 2} e^2 \int d^2x\ {\rm tr}\ 
({1\over \partial _-}
J^+)^2 ,
\eeq
where $J^+ = {i N_f\over 2\pi} h\partial _- h^{\dagger}$. 
In terms of $J= \sqrt \pi J^+$,
the light-cone momentum operators $P^\mu$ take the following simple form
\beq 
P^+={1\over{N_c+N_f}}\int dx^- :J^a(x^-,x^+=0)J^a(x^-,x^+=0): ,
\eeq
namely, the Sugawara form, and
\beq
P^-=-{e^2 \over {2\pi}}\int dx^-:J^a(x^-,x^+=0){1\over{\partial _-
^2}}J^a(x^-,x^+=0): .
\eeq
 Our task will be to solve the eigenvalue equation
\beq
2P^+P^-\state{\psi} = M^2\state{\psi}.
\eeq
We write $P^+$ and $P^-$ in terms of the Fourier transform of $J(x^-)$
defined by $J(p^+)=\int {dx^- \over {\sqrt{2\pi}}} e^{-ip^+x^-} J(x^-,x^+=0)$.
Normal ordering in the expression of $P^+$ and $P^-$ are naturally with respect
to $p$, where $p<0$ denotes a creation operator.
 To simplify the notation we
write from now on $p$ instead of $p^+$. In terms of these variables
the momenta
generators are 
\bea
& & P^+={1\over{N_c +N_f}}\int ^\infty _0dp J^a(-p)J^a(p)     \\
& & P^-={e^2 \over \pi}\int ^\infty _0dp {1\over{p^2}}J^a(-p)J^a(p)
\label{P}
\eea
Recall that the light-cone currents $J^a(p)$ obey a level $N_f$, $SU(N_c)$
affine Lie algebra 
\beq
[J^a(p),J^b(p')]= {1\over 2}N_f\ p\ \delta ^{ab} \delta (p+p')+
 if^{abc} J^c(p+p')
\label{Kac}
\eeq
We can now construct the Hilbert space. The vacuum $\state{0,R}$ is defined
by the annihilation property:
\beq
\forall p>0,\ J(p)\state{0,R}=0
\eeq
Where R is an ``allowed"  representations depending on the level.
Therefore, a physical state in Hilbert space is
$\state{\psi} ={\rm tr}\ J(-p_1)\ldots J(-p_n)\state{0,R}$. Note that
this basis is not orthogonal.

\section{'t Hooft like equation for the two currents wave-function}

Let us restrict ourselves to the 2-currents sector
 of the Hilbert space

\beq
\state{\Phi} = {1\over N_c N_f}\int _0 ^1 dk\ \Phi (k) J^a (-k) J^a (k-1) \state{0},
\eeq

namely to states which are color singlets of two currents with
 total $P^+ =1$ momentum and a distribution of $P^-$ momentum $\Phi(k)$.
 Note that $\Phi$ is a symmetric function
\beq
 \Phi(k) = \Phi(1-k). \label{symm}
\eeq

Our task now is to find the eigenvalue (Schr\"odinger) equation for
 the wavefunction $\Phi(k)$. Let us act with the ``Hamiltonian'' $P^-$
on the state $\state{\Phi}$.

The commutator of $P^-$ with a current $J^b(-k)$ yields the following
 result
\bea
& & [\int _0 ^\infty {dp \over p^2} J^a(-p) J^a(p), J^b(-k)] = \nonumber
 \\
& &
 \left (({1\over 2} N_f - N_c) {1\over k} + N_c {1\over \epsilon} \right) J^b (-k) + \nonumber \\
& & 
\int _k ^\infty dp \left ( {1\over p^2} -
 {1\over (p-k)^2} \right ) if^{abc} J^a(-p) J^c(p-k) 
 + \nonumber \\
& & \int _0 ^k {dp\over p^2} if^{abc} J^c(p-k) J^a(-p). \label{com1}
\eea    

The above expression \eqref{com1} contains 3 terms on the R.H.S. 
The first term contains a single creation operator. The second term
 contains an annihilation current and therefore should be again
 commuted with $J^b (k-1)$. The third term contains two creation
currents and it would lead to a 3-currents state. Namely, the affine Lie
algebra created a higher state. This is a manifestation of the fact that
pair creation is, generically, not suppressed in multi-flavor ${\rm
QCD}_2$, as expected in general in QFT.

 Note that while deriving eq.\eqref{com1} we get an
 ``infinite'' contribution $N_c {1\over \epsilon} J^{b} (-k)$. This
 contribution will be canceled by a counter contribution which comes
 from the regime $p\sim k$ in the first integral on the R.H.S. of
 \eqref{com1}, as below.
 
The commutator of the second term in the R.H.S of \eqref{com1} with
 $J^b (k-1)$ yields

\bea
& & [\int _k ^\infty dp \left ({1\over p^2} -
 {1\over (p-k)^2} \right ) if^{abc} J^a(-p) J^c(p-k), J^b(k-1)]= \label{com2} \\
& & N_c \int _k ^\infty dp \left ( {1\over p^2} -
 {1\over (p-k)^2}\right )
 (J^a (-p) J^a (p-1) - J^a(p-k) J^a(k-p-1)) \nonumber .
\eea

Our results can be summarized by the following set of equations

\bea
\lefteqn{M^2 \state{\Phi} =} \nonumber \\ & & 
  {1\over N_c N_f}\int _0 ^1 dk\ \tilde \Phi (k) J^a (-k) J^a
(k-1) \state{0} + \\ & & 
 {1\over (N_c N_f)^{3\over 2}}\int _0 ^1\ dk\ dp\ dl\ \delta (k+p+l-1) \Psi
(k,p,l) if^{abc} J^a (-k) J^b (-p) J^c (-l) \state{0} \nonumber
\eea
with 
\beq
\Psi(k,p,l) = {2e^2 (N_c N_f) ^{1\over 2} \over \pi }\left (
{\Phi(l) -\Phi(k) \over p^2} \right )
\label{wave3}
\eeq
and
\bea
& &
 \tilde \Phi(k) = {e^2 \over \pi} \left (  
(N_f - N_c ) \left ({1\over k} + {1\over 1-k}\right ) \Phi (k)
+ {2 N_c \over \epsilon} \Phi(k)  \right . \nonumber \\ & &
  \left . - N_c  \int _0 ^{k-\epsilon} dp {\Phi (p) \over (p-k)^2}  
 - N_c  \int _{k+\epsilon} ^1 dp {\Phi (p) \over (p-k)^2}  
          + N_c \left ({1\over k^2} - {1 \over (1-k)^2}
\right ) \int _0 ^k dp\ \Phi (p) \right ). \nonumber
\eea
Ignoring the 3-currents term (see below), we get that
 $\Phi(k)$ obeys the following eigenvalue equation
\bea
& &
 {M^2 \over e^2/ \pi} \Phi(k) = 
(N_f - N_c) \left ({1\over k} + {1\over 1-k}\right ) \Phi (k) \nonumber \\ & &
 - N_c {\cal P} \int _0 ^1 dp {\Phi (p) \over (p-k)^2}  
          + N_c \left ({1\over k^2} - {1 \over (1-k)^2}
\right ) \int _0 ^k dp\ \Phi (p).
\label{tf1}
\eea
For general $N_c$ and $N_f$ discarding the 3-currents term is
unjustified. However, since the length of
 $\Psi$ is $\mid \! \! \Psi (k,p,l) \! \! \mid \sim e^2 (N_c N_f) ^{1\over
2}$, in the limit of large $N_c$ with fixed $e^2 N_c$ and fixed $N_f$, or
 large $N_f$ with fixed $e^2 N_f$ and 
fixed $N_c$, the 3-currents contribution is indeed negligible, as
compared with the 2-currents term, the later being of order 1.
 Note also that while
deriving eq.\eqref{tf1} we assumed that $\int _0 ^1 dp\ \Phi(p) =0$. We
will justify this assumption in the following. Another remark is that
the first
 integral in \eqref{tf1} should be calculated as a principal value
integral. The divergent part of this integral (arising from the regime
$p\sim k$) cancels the previously mentioned infinity.

In order to make contact with the ordinary 't Hooft equation, it
 is useful to integrate equation \eqref{tf1} with respect to $k$
and rewrite the equation in terms of $\varphi (k) \equiv \int _0 ^k  dp\ \Phi (p)$.
\bea
& &
 {M^2 \over e^2/ \pi} \varphi(k) = 
(N_f - N_c) \left ({1\over k} + {1\over 1-k}\right ) \varphi (k) \nonumber \\ & &
 - N_c {\cal P}\int _0 ^1 dp {\varphi (p) \over (p-k)^2}  
 + N_f \int _0 ^k dp {\varphi (p) \over p^2} + N_f \int _k ^1 dp
{\varphi(p) \over (1-p)^2}
\label{tf2}
\eea
The derivation goes as follows. First, integrating eq.\eqref{symm} we
get $\varphi (k) = - \varphi(1-k) +{\rm const. }$
 Taking $\varphi(1)=0$ we get
\beq
\varphi (k) = - \varphi(1-k). \label{tw}
\eeq
Now $\varphi(1)=0$ implies $\int _0 ^1 dk \Phi(k)=0$, which was
our assumption above. Then, differentiating \eqref{tf2} we do get
\eqref{tf1}, and by \eqref{tw} we also get that there is no extra
integration constant. 

We would like to comment on the issue of the Hermiticity of the
'Hamiltonian' $M^2$. Naively, it seems that $M^2$ is not Hermitian
with respect to the scalar product 
$< \psi | \varphi> = \int _0 ^1 dk \psi ^\star (k) \varphi(k)$, since the
Hermitian conjugate of \eqref{tf2} is
\bea
& &
 \left ( {M^2 \over e^2/ \pi} \right ) ^\dagger \varphi(k) = 
(N_f - N_c) \left ({1\over k} + {1\over 1-k}\right ) \varphi (k)
 \nonumber \\ & &
 - N_c {\cal P}\int _0 ^1 dp {\varphi (p) \over (p-k)^2}  
 - N_f {1\over k^2} \int _0 ^k dp \varphi (p) - N_f {1\over (1-k)^2} 
\int _k ^1 dp \varphi(p)
\label{tf3}
\eea
However, as we shall see in the next section, the numerical solution
yields real eigenvalues and eigenfunctions. Therefore, at least on the
subspace which is spanned by the eigenfunctions, namely real functions
which are zero at $k=0,1$ and anti-symmetric with respect to
$k={1\over 2}$, the operator $M^2$ is Hermitian. Note that \eqref{tf3}
is ``more regular'' than \eqref{tf2}, as in \eqref{tf2} it is
$\varphi(p)/p^2$ that appears in the integration from zero.

Equation \eqref{tf2} is similar to 't Hooft equation for a massive single
flavor large $N_c$ ${\rm QCD}_2$, with $m^2 = {e^2 N_f \over \pi}$. It differs 
from 't Hooft's equation by having two additional terms (two last terms
in \eqref{tf2}). It suggests that the dynamics which governs the
lowest state of the multi-flavor model is given, approximately,
 by a model of massive 
``glueball'' with an $SU(N_c)$ gauge interaction and additional terms which
 are proportional to $N_f$.

Before we present our solution of \eqref{tf2} it is important to note 
that it is only an approximated solution. We neglected the 3-currents
 state with, a-priori, no justification. We shall see, however, that
the restriction to the truncated 2-currents sector
 is an excellent approximation for the lowest massive meson.

\section{The spectrum - numerical results} 
 
The most convenient way to solve \eqref{tf2} is to expand $\varphi(k)$
in the following basis (see, however \cite{abe}, a different interesting
choice of basis)
\beq
\varphi(k) = \sum _{i=0} ^\infty A_i (k-{1\over 2})\left (k(1-k)\right
)^{\beta +i} \label{basis}
\eeq
The value of $\beta$ chosen such that the
Hamiltonian will not be singular near $k\rightarrow 0$ (or $k\rightarrow
1$)\cite{thooft},\cite{sande}. This consideration
 leads to the following equation
\beq
({N_f \over N_c} -1) - {N_f/N_c \over \beta +1 } + \beta \pi \cot
\beta \pi = 0 \label{beta}.
\eeq
This comes from eq.\eqref{tf3}. Had we started with \eqref{tf2}, it
would have been $-\beta$ replacing $\beta$ in \eqref{beta}, and
constrained to $\beta$ larger than 1.
Upon truncating the infinite sum in \eqref{basis} to a finite sum,
the eigenvalue problem reduces to a diagonalization of a matrix.
So, the problem can be reformulated as follows
\beq
 \lambda N_{ij} A_j = H_{ij} A_j,
\eeq
with 
\beq
 N_{ij} = \int _0 ^1 dk (k-{1\over 2})^2 \left (k(1-k)\right )
^{2\beta + i+j},
\eeq
and
\bea
& &
H_{ij} = ({N_f\over N_c} -1)\int _0 ^1 dk (k-{1\over 2})^2 \left (
k(1-k) \right ) ^{2\beta +i+j-1} \nonumber \\ 
& & - {N_f\over N_c} \int _0 ^1 dk
(k-{1\over 2}) \left ( k(1-k) \right ) ^{\beta +i} {1\over k^2} 
\int _0 ^ k (p-{1\over 2}) \left ( p(1-p) \right) ^{\beta +j}
\nonumber \\
& & - {N_f\over N_c} \int _0 ^1 dk
(k-{1\over 2}) \left ( k(1-k) \right ) ^{\beta +i} {1\over (1-k)^2} 
\int _k ^1 (p-{1\over 2}) \left ( p(1-p) \right) ^{\beta +j}
\nonumber \\
& &  - \int _0 ^1 dk dp { (k-{1\over 2})\left (k(1-k) \right ) ^{\beta
+i} (p-{1\over 2})\left (p(1-p) \right ) ^{\beta+j} \over (k-p)^2}
\eea
Hence
\beq
N_{ij} ={ B(2\beta +i+j+2,2\beta +i+j+2) \over {2( 2\beta +i+j+1)}} ,
\eeq
and
\bea
& &
H_{ij} = ({N_f\over N_c} -1) {B(2\beta +i+j+1,2\beta +i+j+1)
 \over { 2(2\beta +i+j) }}
\nonumber \\
& &
 - {N_f\over N_c} {B(2\beta +i+j+1,2\beta +i+j+1)
 \over 2( 2\beta +i+j ) (\beta +j+1 ) } \nonumber \\
& & + {(\beta +i)(\beta +j)
B(\beta +i,\beta+i) B(\beta +j,\beta +j) \over
8(2\beta+i+j)(2\beta +i+j+1) } , 
\eea
where $B(x,y)$ is the Beta function
\beq
B(x,y) = {\Gamma (x) \Gamma(y) \over \Gamma (x+y)}. 
\eeq
In practice, the process converges rapidly and a $5\times 5$ matrix
 yields the 'continuum' results.
 
The lowest eigenvalues of \eqref{tf2} as a function of the ratio
${N_f \over N_c}$ are listed in table 1 below (see also figure 1).
Note that by $\beta =0, N_f/N_c =0$ we mean the limit $\beta
\rightarrow 0,N_f/N_c \rightarrow 0$.
\begin{table}[H]
\centerline{
\begin{tabular}{|c|c|c|}
\hline
$\beta$ & $N_f / N_c$ & $M^2$
\\ 
\hline
0.0000 &  0   & 5.88 \\
0.0573 &  0.2 & 6.91 \\
0.1088 &  0.4 & 7.91 \\
0.1552 &  0.6 & 8.91 \\
0.1978 &  0.8 & 9.89 \\
0.2366 &  1.0 & 10.86 \\
0.2725 &  1.2 & 11.83 \\
0.3050 &  1.4 & 12.77 \\
0.3360 &  1.6 & 13.73 \\
0.3645 &  1.8 & 14.67 \\
\hline
\end{tabular}
}
\caption{The mass of the lowest massive meson, in units of ${e^2 N_c
\over \pi}$, as a function of $N_f/N_c$ and $\beta$.}
\end{table}

These values are in excellent agreement with recent DLCQ calculations.
For comparison see \cite{GHK}, \cite{AP} and especially \cite{uwe} for
a recent work. The typical error is less than 0.1\% !

An interesting observation is that the eigenvalues depends
linearly on $N_f$ (see figure 1). The dependence is 
\beq
M^2 = {e^2 N_c
\over \pi} (5.88 + 5 {N_f\over N_c}) .
\eeq
We do not have a good
understanding of this observation. It is not clear why the lowest
eigenvalue sits on a straight line. It is not clear even why, as a
eigenvalue equation \eqref{tf2} exhibits such a behavior.    

In the following sections we will consider some special cases.

\subsection{$N_f=1$, The 't Hooft Model}

The limit $N_c \rightarrow \infty$ with $e^2 N_c$ fixed and $N_f \ll
N_c$ corresponds to the well known 't Hooft model. In this limit ${\rm QCD}_2$
was solved exactly long time ago by 't Hooft\cite{thooft}.
Let us see how our approach coincides with the fermionic basis in this
case. In the limit $N_f \ll N_c$ we can neglect terms which are
proportional to $N_f$. Equation \eqref{tf2} (or \eqref{tf3})
takes the following form
\beq
 {M^2 \over e^2/ \pi} \varphi(k) = 
 - N_c \left ({1\over k} + {1\over 1-k}\right ) \varphi (k) \nonumber 
 - N_c {\cal P}\int _0 ^1 dp {\varphi (p) \over (p-k)^2},
\label{tf}
\eeq
which is just 't Hooft equation for the massless case. Note that
\eqref{tf} is {\em exact}, since in the small $N_f$ limit the
3-currents state is suppressed by $ N_c ^{-{1\over 2}}$ with respect to the 
2-currents state and therefore we can neglect it.

Since the
wavefunction $\varphi(k)$ is anti-symmetric, we
will recover only the odd states in the spectrum of ${\rm QCD}_2$ (the even 
states can be recovered by considering other sectors of the Hilbert
 space which decouple from the 2-currents state).

Though equation \eqref{tf} is formally the same as 't Hooft
equation, the interpretation of $\varphi(k)$ should be different.
$\varphi(k)$ is the integral of the function $\Phi(k)$ which
corresponds to 2-currents state, namely to a mixture of 4-fermions and
2-fermions. What is the relation between the states that we find here
 and the mesons in 't Hooft's model ?

In order to answer this question let us expand the currents in terms
 of fermions. It is useful to denote the current in double index notation
\beq
J^{a} (k) \rightarrow J^i _j (k) = \int _{-\infty} ^{\infty} dq\ 
\left ( \bar \Psi ^i(q) \Psi _j(k-q) - {1\over N_c} \delta ^i _j 
        \bar \Psi ^k(q) \Psi _k(k-q) \right )
\eeq
We do not bother about normal ordering, as no problem for $k$
 non zero, and we have to treat the $k=0$ part in a limiting way.
The state $\state{\Phi}$ can be written as
\bea
\lefteqn{\state{\Phi} = {1\over 2N_c} \int _0 ^1 dk\ \Phi(k) J^i _j (-k) J^j _i (k-1)
 \state{0} =}  \label{ferm} \\
& &
  {1\over 2N_c} \int _0 ^1 dk\ \Phi (k) \int _{-\infty}
 ^{\infty} dq\ dp\ 
\left ( \bar \Psi ^i (-q) \Psi _j (-k+q) - {1\over N_c} \delta ^i _j
        \bar \Psi ^k (-q) \Psi _k (-k+q) \right ) \times \nonumber \\
& & 
\left ( \bar \Psi ^j (-p) \Psi _i (k+p-1) - {1\over N_c} \delta ^j _i
        \bar \Psi ^k (-p) \Psi _k (k+p-1) \right) \state{0}. \nonumber
\eea
Note that the above expression \eqref{ferm} contains annihilation and creation
 fermionic operators. Written in terms of creation operators only,
 \eqref{ferm} reads
\bea
& & 
\state{\Phi} = {1\over 2N_c} \int _0 ^1 dk \int _0 ^k dq \int _0
 ^{1-k} dp\ \Phi(k) \bar \Psi ^i (-q) \Psi _j (-k+q) \bar \Psi ^j (-p)
 \Psi _i (k+p-1) \state{0}  \nonumber \\
& &
- {1\over 2N_c^2 } \int _0 ^1 dk \int _0 ^k dq \int _0 ^{1-k} dp\ \Phi(k) \bar \Psi ^i (-q) \Psi _i (-k+q) \bar \Psi ^j (-p)
 \Psi _j (k+p-1) \state{0}  \nonumber \\
& &
- (1-{1\over N_c^2})
\int _0 ^1 dk \int _0 ^k dq\ \Phi(k) \bar \Psi ^i (-q) \Psi _i (q-1)
\state{0} \label{meson}
\eea
The last term in \eqref{meson} corresponds to a meson. It can be
 written also as
\beq
 \int _0 ^1 dq \int _q ^1 dk\ \Phi(k)   \bar \Psi ^i (-q) \Psi _i (q-1)
\state{0} = - \int _0 ^1 dq\ \varphi(q)  \bar \Psi ^i (-q) \Psi _i (q-1)
\state{0}
\eeq
 which is exactly the 't Hooft meson. We conclude that the 2 currents
 state has an overlap with the 't Hooft meson and this is why
 \eqref{tf2} reproduces exactly the (odd part of the) spectrum of the
 't Hooft model.   

\subsection{Large $N_f \gg N_c$ limit}

In the limit $N_f \gg N_c$, with $e^2 N_f$ fixed, the truncation to
2-currents state should again predict exact results. The reason
 is that the 3-currents state is suppressed by $N_f ^{-{1\over
2}}$ with respect to the 2-currents state.

In this limit eq.\eqref{tf1} takes the form 
\beq
M^2 = {e^2 N_f \over \pi} \left ( {1\over k} + {1\over 1-k} \right ).
\label{NF}
\eeq

It describes a continuum of states with
 masses above $2m$, where $m^2 ={e^2 N_f \over \pi}$.
The interpretation is clear: in this limit the spectrum of the 
theory reduces to a single non-interacting meson (or ``currentball'')
with mass $m$. This phenomena
 was already observed in \cite{AS} and in \cite{AFST} by using a 
different approach. 

\subsection{$N_f = N_c$, The Adjoint Fermions Model}

The case $N_f=N_c$ is the most interesting one. It was shown 
 that the massive spectrum of this model is equivalent to the massive 
spectrum of a model with a single adjoint fermion, due to
 'universality' \cite{KS}. Since this model is not exactly solvable, 
it is interesting to see how our approach reproduces, almost
 accurately, previous numerical results.

The mass of the lowest massive meson, predicted by \eqref{tf2}, is
$M^2 = 10.86 \times {e^2 N_c \over \pi}$. For comparison,
the recent values reported in the literature are $M^2=10.8$ \cite{GHK}
and $M^2=10.84$ \cite{uwe}, in units of ${e^2 N_c \over \pi}$.

This agreement is very surprising. In the regime $N_f \sim N_c$, the 3-currents
 state is not suppressed by factors of color or flavor with respect to
 to the 2-currents state. Why, thus, is our approach so successful ?
The reason seems to be that as in the fermionic
 basis\cite{BDK},
 the lowest massive state is an almost pure 2 currents state. However,
 the present approach is much more successful than the fermionic basis,
 where the prediction for the mass of the lowest massive boson of the 
adjoint model is twice as much as the lowest massive boson of 't Hooft
 model. It seems that the 'correct' underlying degrees of freedom
are currents and not fermions, as predicted by the authors of \cite{KS}. 

\section{Summary}
In this work we used a description of massless ${\rm QCD}_2$
 in terms of currents.
With this basis we wrote down  a 't Hooft like equation
\eqref{tf1} for the wave function of the two currents states. 
 
The equation interpolates smoothly between the description of a single  
flavor model with large $N_c$ ('t Hooft model), the adjoint fermions model $N_f=N_c$ and
the large $N_f$ model. 
The equation is derived by using an a-priori  unjustified suppression of   
the three currents coupling. Nevertheless, we observe an excellent
agreement with the DLCQ results for the first excited state. For
higher excited states the agreement deteriorates and it is of the
order of 20\%.

The accuracy of the results for the first excited state,
which implies that for this state the truncation of the ``pair creation
terms" is harmless,   deserves further  investigation.

\Acknowledgements
A.A. thanks the department of particle physics at
the Weizmann institute of science
for the warm hospitality while part of this work was done. 
The research of J.S was supported in part by
 the US--Israeli Binational Science Foundation,
 the Israeli Science Foundation,
 the German--Israeli Foundation for Scientific Research (GIF)
 and the Einstein Center for theoretical Physics at the Weizmann Institute.
A.A would
also like to thank B. Van de Sande for help with the numerical
solution
 of 't Hooft equation.
The work of A.A. is supported in part by EEC under TMR contract
ERBFMRX-CT96-0090.

\newpage

 \begin{figure}[H]
  \begin{center}
\mbox{\kern-0.5cm
\epsfig{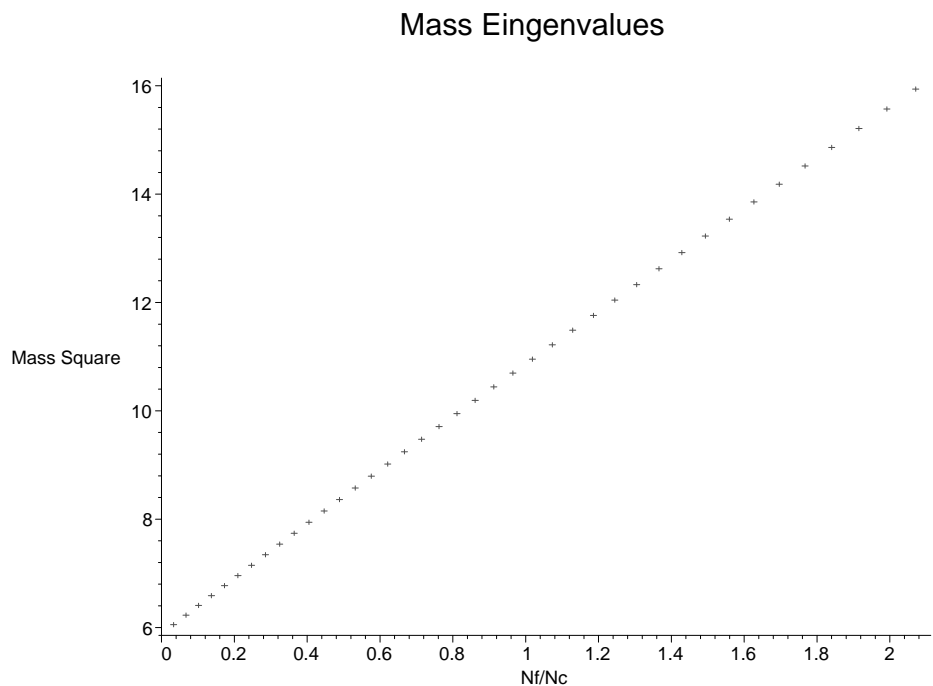}}
\label{rules2}
  \end{center}
\caption{The mass of the lowest massive meson, in units of ${e^2 N_c
\over \pi}$, as a function of $N_f/N_c$.}
\end{figure}

\end{document}